\documentclass{PoS}

\title{Transverse-spin gluon distribution function}

 \ShortTitle{Transverse-spin gluon distribution function}

 \author{\speaker{Kazuhiro TANAKA}%
\\
          Department of Physics, Juntendo University, Inzai, Chiba 270-1695, Japan\\
and\\
J-PARC Branch, KEK Theory Center, Institute of Particle and Nuclear Studies, High Energy Accelerator Research Organization (KEK), 203-1, Shirakata, Tokai, Ibaraki, 319-1106, Japan\\
         E-mail: \email{tanakak@sakura.juntendo.ac.jp}}

\abstract{We introduce the spin-operator representation for the gluon as well as quark distribution functions
as nucleon matrix element of the gauge-invariant bilocal light-cone operators in QCD.
To identify the relevant spin operators for quarks and gluons in a unified manner,
we rely on the transformation properties of the quark and gluon fields in the coordinate space
under the action of the generator of the Lorentz group.
In particular, this approach allows us 
to define the transverse-spin gluon distribution function
$G_T(x)$, which is the genuine counterpart of the transverse-spin quark distribution function $g_T(x)$ relevant to the transverse-spin structure function $g_2(x, Q^2)$ in the deep inelastic scattering.
We show that $G_T(x)$ is given by the sum of the chromoelectric and chromomagnetic correlators 
associated with helicity-flip by one unit, and the treatment of the latter correlator
completes the classification of the collinear parton distribution functions up to twist three.
We show that $G_T(x)$ receives the three-gluon and quark-gluon correlation effects and
discuss the operator product expansion for $G_T(x)$.
We also discuss the relevance of the first moment of $G_T(x)$ for the partonic decomposition of the transverse nucleon spin.}

\FullConference{XXII. International Workshop on Deep-Inelastic Scattering and Related Subjects,\\
		28 April - 2 May 2014\\
		Warsaw, Poland}

\usepackage{bm}
\newcommand{\Slash}[1]{{\ooalign{\hfil/\hfil\crcr$#1$}}}
\begin{document}
In hard processes like deep inelastic scattering (DIS), Drell-Yan process, etc., the cross sections are given
by the hard partonic scattering combined with the parton distribution functions (PDFs),
as the factorization formulas.
The PDFs are given by the Fourier transformation of the bilocal operators 
instantaneous for a relevant light-cone direction, e.g., $z^+$.
For the hard processes with one hard scale, the transverse degrees of freedom can be also
integrated out in the factorization formulas, leaving the collinear PDFs as the one-dimensional light-cone
Fourier transform; e.g., $\sim \int {d{z^ - }{e^{i(x{P^ + }){z^ - }}}} \left\langle P S \right|{\psi ^\dag }(0)\psi (
{z^ - }
)\left| P S\right\rangle$ for the quark distributions in the nucleon with momentum $P$ and spin $S$ ($P^2=-S^2 = M^2$);
here and in the following, we omit the Wilson line
operator in-between the constituent fields.
Inserting the various gamma matrices 
in-between the quark and antiquark fields gives
a complete set of the quark distribution functions~\cite{Collins:1981uw,Manohar:1990kr,Jaffe:1991ra} (see also \cite{Kodaira:1998jn}):
\begin{equation}
\int {\frac{{d\lambda }}{{2\pi }}} {e^{i\lambda x}}\langle P{\rm{ }}S|\bar \psi_\beta (0)\psi_\alpha (\lambda n)|P{\rm{ }}S\rangle
 = \frac{1}{2}\left[q(x) \Slash{P}
+ \Delta q(x)\left( {S\cdot n} \right)  \gamma_5\Slash{P}
+ {g_T}(x)  \gamma_5\Slash{S}_\bot \right]_{\alpha \beta} \ ,
\label{qpdf}
\end{equation}
with a light-like vector ${n^\mu } =g^\mu_- /P^+$. 
Here and in the following, we treat the distribution functions up to twist three:
the twist-two density and helicity quark distributions, $q(x)$ and $\Delta q(x)$,
and the twist-three transverse-spin quark distribution $g_T(x)$. These chiral-even
distributions are relevant in the following discussions; in (\ref{qpdf}),
we omitted the terms associated with chiral-odd distributions.

The gluon distribution functions may be introduced~\cite{Collins:1981uw,Manohar:1990kr,Ji:1992eu,Hatta:2012jm} by replacing the quark fields $\psi$ with
the gluon fields $A^\mu$,
in particular, 
by the formal replacement
$\psi \rightarrow n^-F^{ + \nu}\equiv F^{ n \nu}$ with $F^{\mu \nu}$ the gluon field strength tensor, 
as motivated by 
$F^{+\nu}=\partial^+ A^\nu$ in the $A^+=0$ gauge:
\begin{equation}
\frac{- 1}{x}\int {\frac{{d\lambda }}{{2\pi }}} {e^{i\lambda x}}\langle P{\rm{ }}S|{F^{ n \nu }}(0){F^{ n \sigma }}(\lambda n)|P{\rm{ }}S\rangle
= \frac{1}{2}\left[  G(x){g_\bot^{\nu \sigma }}  + \Delta G(x)i{\epsilon^{\nu \sigma Pn}}\left( {S\cdot n} \right) + 2{G_{3E}}(x)i{\epsilon^{\nu \sigma \alpha n}}{S_{ \bot \alpha }} \right] ,
\label{gpdf}
\end{equation}
where 
$g_\bot^{\nu \sigma }=g^{\nu\sigma}-P^\nu n^\sigma -P^\sigma n^\nu$, 
and $\epsilon^{\nu \sigma Pn}\equiv  \epsilon^{\nu \sigma \alpha \beta}P_\alpha n_\beta$ with
the Levi-Civita tensor of
$\epsilon_{0123}=1$.
The RHS is analogous to (\ref{qpdf}),
with the twist-two density and helicity distributions, $G(x)$ and $\Delta G(x)$,
and the twist-three gluon distribution $G_{3E}(x)$,
which was denoted as $G_3(x)$ (${\cal G}_{3T}(x)$) in \cite{Ji:1992eu}
(\cite{Kodaira:1998jn,Hatta:2012jm}).

Now, in order to go beyond the conventional treatment (\ref{gpdf}),
we consider the Lorentz tensor decomposition of the most general gluonic correlator and find, to the 
twist-three accuracy,
\begin{eqnarray}
\lefteqn{ - \frac{1}{x}\int {\frac{{d\lambda }}{{2\pi }}} {e^{i\lambda x}}\langle P{\rm{ }}S|{F^{\mu \nu }}(0){F^{\xi \sigma }}(\lambda n)|P{\rm{ }}S\rangle }
\nonumber\\
 &&= \frac{1}{2} G(x)\left( {{P^\mu }{P^\xi }g_ \bot ^{\nu \sigma } - {P^\nu }{P^\xi }g_ \bot ^{\mu \sigma } - {P^\mu }{P^\sigma }g_ \bot ^{\nu \xi } + {P^\nu }{P^\sigma }g_ \bot ^{\mu \xi }} \right)
\nonumber\\
&& 
+ \frac{1}{2}  \Delta G(x)i\left( {S\cdot n} \right)P_\alpha\left( {{P^\mu }{\epsilon^{\nu \xi \sigma \alpha }} - {P^\nu }{\epsilon^{\mu\xi \sigma \alpha }}} \right)
\nonumber\\
&& + {G_{3E}}(x)i{S_{ \bot \alpha }}\left( {{P^\mu }{\epsilon^{\nu \xi \sigma \alpha }} - {P^\nu }{\epsilon^{\mu \xi \sigma \alpha }}} \right)
 + {G_{3H}}(x)i P_\alpha \left( {S_ \bot ^\mu {\epsilon^{\nu\xi  \sigma \alpha }} - S_ \bot ^\nu {\epsilon^{\mu \xi \sigma \alpha }}} \right) \ ,
\label{ff}
\end{eqnarray}
consistent with
PT-invariance and hermiticity~\cite{kt}.
When contracted with $n_\mu n_\xi$, the tensor structure with ${G_{3H}}(x)$
vanishes, while the other terms reproduce the formula~(\ref{gpdf}).
The new distribution ${G_{3H}}(x)$ is of twist three and is associated with the
transverse spin of the nucleon.
This new contribution seems to violate the correspondence between the
quark and gluon cases, as suggested in (\ref{qpdf}) and (\ref{gpdf}), and thus it is desirable to clarify the physical meaning of $G_{3H}(x)$ as well as of $G_{3E}(x)$.

For this purpose, we first note that (\ref{qpdf}) with $P=(P^0,0,0,P^3)$, $P^\pm=\frac{P^0\pm P^3}{\sqrt{2}}$ can be recast as
\begin{eqnarray}
\int {\frac{{d\lambda }}{{2\pi }}} {e^{i\lambda x}}\langle P{\rm{ }}S|\psi_\beta^\dag (0)\psi_\alpha (\lambda n)|P{\rm{ }}S\rangle
 &&\!\!\!\!= \frac{1}{\sqrt{2}}\left[  q(x)  P^+{\cal P}_{(+)}
+  2\Delta q(x)\left( {S\cdot n} \right) P^+  {\cal P}_{(+)} \hat{s}^3
\right.
\nonumber\\
&&\!\!\!\!\!\!\!\!\!\!\!\!+\left. \sqrt{2}\ {g_T}(x)  
\left( {\cal P}_{(-)} {\bm S}_\bot \cdot \hat{\bm{s}} {\cal P}_{(+)}
+{\cal P}_{(+)} {\bm S}_\bot \cdot \hat{\bm{s}} {\cal P}_{(-)}\right)
 \right]_{\alpha \beta} \ ,
\label{qpdfr}
\end{eqnarray}
where ${\cal P}_{(\pm)}=\frac{1}{2}\gamma^\mp \gamma^\pm$ 
are the projection operators 
with 
${\cal P}_{(+)}{\cal P}_{(-)}=0$,
${\cal P}_{(+)}+{\cal P}_{(-)}=1$,
such that ${\cal P}_{(+)}\psi$ and ${\cal P}_{(-)}\psi$
are the ``good'' and ``bad'' components 
of the quark fields in the light-cone quantization formalism,
and $\hat{s}^i= \frac{1}{4}\eta^{i j k}\sigma^{jk}$ 
are the quark spin operators.\footnote{The repeated indices should be understood as summed over. We use Latin letters $i,j=1,2,3$ for three-dimensional space indices and also use the three-dimensional totally antisymmetric tensor $\eta^{ijk}\equiv \epsilon^{ijk0}$, $\eta^{123}=+1$.} From (\ref{qpdfr}), we find
$(n^-/\sqrt{2})\int {\frac{{d\lambda}}{{2\pi }}} {e^{i\lambda x}}\langle P| \psi^\dagger(0)
\psi (\lambda n)|P\rangle= q(x)$,
$\sqrt{2}n^-\int {\frac{{d\lambda}}{{2\pi }}} {e^{i\lambda x}}\langle P{\rm{ }}S_\parallel| \psi^\dagger(0)\hat{s}^3
\psi (\lambda n)|P{\rm{ }}S_\parallel\rangle
= \Delta q(x)$, and
$(1/M)\int {\frac{{d\lambda}}{{2\pi }}} {e^{i\lambda x}}\langle P{\rm{ }}S_\bot| \psi^\dagger(0)\hat{s}^\bot
\psi (\lambda n)|P{\rm{ }}S_\bot\rangle
= g_T(x)$,
up to the twist-four corrections suppressed for the fast-moving nucleon by powers of
$M/P^+$.
These representations demonstrate that the ``quantum mechanical'' expectation values
in the spin space, $\psi^\dag  \psi$,  $\psi^\dag \hat{s}^i \psi$, are relevant for 
the quark distributions of low twist. 
We note that the twist-three distribution $g_T(x)$ is a complicated object, 
because 
${\cal P}_{(-)}\psi=-(2n\cdot\partial)^{-1} \Slash{n}
{\Slash{D}_\bot }{{\cal P}_{(+)}\psi}$,
using the QCD equations of motion with $A^+=0$, and the good components ${\cal P}_{(+)}\psi$ are the independent degrees of freedom
in the light-cone quantization:
$q(x)$, $\Delta q(x)$ are 
literally the distributions, while $g_T(x)$ 
represents the quark-gluon three-body correlations.

The spin operator for a particle can be identified in the $x\rightarrow 0$ limit of 
the transformation law of the corresponding field $\Phi(x)$ under the action of the generator ${\cal M}^{\mu \nu}$ of the Lorentz group:
\begin{equation}
\left[ {{{\cal M}^{\mu \nu }},\Phi (x)} \right] =  - \left( {i\left( {{x^\mu }{\partial ^\nu } - {x^\nu }{\partial ^\mu }} \right)
 + \Sigma^{\mu \nu}} \right)\Phi (x)\ ,
\label{trg}
\end{equation}
and this guarantees
the usual $SU(2)$ algebra, 
$\left[\hat{s}^i, \hat{s}^j \right] = i\eta^{ijk}\hat{s}^k $,
obeyed by $\hat{s}^i \equiv \frac{1}{2}\eta^{i j k}\Sigma^{jk}$.
Indeed, we obtain $\Sigma^{\mu \nu}=\frac{1}{2}{\sigma ^{\mu \nu }}$ for $\Phi=\psi$.
Similarly,
manifestly gauge-covariant form of the transformation law for the gluon is 
given by (\ref{trg}) with 
$\Phi(x) = {F^{\alpha \beta }}(x)$,
${\Sigma ^{\mu \nu }}{F^{\alpha \beta }} = i( {g^{\mu \alpha }}{F^{\nu \beta }} - {g^{\nu \alpha }}{F^{\mu \beta }} + {g^{\mu \beta }}{F^{\alpha \nu}} - {g^{\nu \beta }}{F^{\alpha \mu}} )$,
so that
$\hat{s}^i \equiv \frac{1}{2}\eta^{i j k}\Sigma^{jk}$ give the spin operators for a spin-1 particle.
This fact allows us to extend the above spin-operator representation of the PDFs to the gluon case.
The corresponding ``quantum mechanical'' expectation values
in the spin space should be given by 
$(F^{\alpha \beta })^{\dag} {F^{\alpha \beta}}$, 
$(F^{\alpha \beta })^{\dag} \hat{s}^i {F^{\alpha \beta}}$; note,
$(F^{\alpha \beta })^{\dag} {F^{\alpha \beta}}
\neq (F^{\alpha \beta })^\dag {F_{\alpha \beta}}$, 
where 
$(F^{\alpha \beta })^\dag {F_{\alpha \beta}}=F^{\alpha \beta } {F_{\alpha \beta}}$ is 
a Lorentz invariant contraction.
We find,
$(F^{\alpha \beta })^\dag{F^{\alpha \beta }}=-2F^{ + \beta }{F^ + }_\beta + \cdots$,
and $(F^{\alpha \beta })^\dag \hat{s}^3{F^{\alpha \beta }}=2i{F^{ + \beta }}
{\widetilde F^ + }_{\ \  \beta}+\cdots$, 
so that
\begin{eqnarray}
\frac{(n^-)^2}{2x}\int {\frac{{d\lambda }}{{2\pi }}} {e^{i\lambda x}}\langle P|
\left(F^{\alpha \beta } (0)\right)^\dag{F^{\alpha \beta }}(\lambda n)|P\rangle
&=& G(x) 
\ ,
\label{gspin}\\
\frac{(n^-)^2}{2x}\int {\frac{{d\lambda }}{{2\pi }}} {e^{i\lambda x}}\langle P{\rm{ }}S_\parallel|
\left(F^{\alpha \beta } (0)\right)^\dag
{\hat s^3}{F^{\alpha \beta }}(\lambda n)|P{\rm{ }}S_\parallel\rangle
&=& \Delta G(x)
\ ,
\label{gspinl}
\end{eqnarray}
up to twist-four corrections, suppressed as
$\sim (M/P^+)^2$ for the fast-moving nucleon, and 
similarly, 
\begin{eqnarray}
&&
\!\!\!\!\!\!\!\!
\frac{n^-}{2\sqrt{2}M x}\int {\frac{{d\lambda }}{{2\pi }}} {e^{i\lambda x}}\langle P{\rm{ }}S_\bot|
\left(F^{\alpha \beta } (0)\right)^\dag
\hat{s}^\bot {F^{\alpha \beta }}(\lambda n)|P{\rm{ }}S_\bot\rangle
= 
\frac{-n^-}{2M x}\int {\frac{{d\lambda }}{{2\pi }}} {e^{i\lambda x}}
\langle P{\rm{ }}S_\bot|i\widetilde{F}^{ + \bot }(0)F^{+-} (\lambda n)
\nonumber\\
&& 
\;\;\;\;\;\;\;\;\;\;\;\;\;\;\;\;\;\;\;\;\;\;\;\;\;\;\;\;\;\;\;\;\;\;\;\;\;
\;\;\;\;\;\;\;\;\;\;\;\;\;\;\;\;\;\;
+
i{F^{ + \bot }}(0) F^{12} (\lambda n)+ {\rm h.c.}
| P{\rm{ }}S_\bot\rangle 
= G_{3E}(x)+G_{3H}(x)\equiv G_T(x)
\ ,
\label{gspint}
\end{eqnarray}
with ``h.c.'' denoting the hermitian conjugate of the preceding terms, where 
the contributions of operators involving $F^{+-}$ 
can be expressed by $G_{3E}$ using (\ref{gpdf}), while
those involving $F^{12}$ require
the new distribution $G_{3H}$ of (\ref{ff}).
We denote the resulting sum
$G_{3E}(x) + G_{3H}(x)$ as $G_{T}(x)$.
The QCD equations of motion in the $A^+=0$ gauge give
$F^{ +  - } = \frac{1}{{{n\cdot\partial }}}\left( {{D_{ \bot j}}{F^{j n }} + g\bar \psi {t^a}\Slash{n}\psi {t^a}} \right)$,
which
shows that $G_{3E}(x)$ is related to the three-gluon correlations
and the quark-gluon correlations. Similarly,
${F^{12}} = {\partial ^1}{A^2} - {\partial ^2}{A^1} - ig[{A^1},{A^2}]$
shows that $G_{3H}(x)$ is also related to the
three-gluon correlations.
Therefore, $G_T(x)$ is a complicated three-body object 
in contrast to 
$G(x)$, $\Delta G(x)$.
In particular, comparing (\ref{gspin})-(\ref{gspint}) with (\ref{qpdfr}), $G_T(x)$ 
is the genuine gluonic analogue of the
twist-three transverse-spin quark distribution $g_T(x)$, so that 
be identified as the transverse-spin gluon distribution.

As is well-known, the helicity structures 
in (\ref{qpdf}) can be revealed by 
decomposing the quark field $\psi$ into the helicity-up and -down 
components, $\psi=\psi_\uparrow +\psi_\downarrow$ with $\psi_\uparrow = \frac{1}{2}(1+ \sigma^{12})\psi$, $\psi_\downarrow = \frac{1}{2}(1-\sigma^{12})\psi$:
$q(x)$ and $\Delta q(x)$
pick up the combinations,
$\psi_\uparrow^\dag  \psi _ \uparrow  + \psi _ \downarrow ^\dag 
\psi _ \downarrow $ and 
$\psi _ \uparrow ^\dag  \psi _ \uparrow  - \psi _ \downarrow ^\dag 
\psi _ \downarrow $, respectively,
in the bilinear operator in the LHS of (\ref{qpdfr}),
corresponding to the density and helicity distributions;
on the other hand, the operator $\bm{S}_\bot \cdot \hat{\bm{s}}$ for $g_T(x)$ 
picks up the combinations, $\psi _ \uparrow ^\dag {\psi _ \downarrow }$,
$\psi _ \downarrow ^\dag {\psi _ \uparrow }$,
demonstrating that the transverse-spin distribution corresponds to helicity-flip
by one unit.

To show the helicity structures of the gluon distributions,
we introduce the right- and left-handed circular polarization
vectors, $\epsilon_R^\mu=(0, - 1, - i,0)/\sqrt 2$,
$\epsilon_L^\mu=(0, + 1, - i,0)/\sqrt 2$, so that 
the contraction with these vectors represents the helicity $\pm 1$ states.
The unit matrix in-between the gluon field strength 
tensors in the LHS of (\ref{gspin})
picks up the combination, 
$(F^{ + R})^\dag {F^{ + R}} + (F^{ + L})^\dag {F^{ + L}}$,
and the operator $\hat{s}^3$ of (\ref{gspinl})
picks up, $(F^{ + R})^\dag {F^{ + R}} - (F^{ + L})^\dag {F^{ + L}}$,
clarifying that $G(x)$ and $\Delta G(x)$ are indeed the density and 
helicity distributions~\cite{Manohar:1990kr}.
For (\ref{gspint}),
the bilocal operator ${\widetilde F^{ +  \bot }}{F^{ +  - }}$
relevant to $G_{3E}(x)$ is expressed by
$(F^{ + R})^\dag E^3$, $(F^{ + L})^\dag E^3$
with $E^3=F^{30}=F^{+-}$ being
the third component of the chromoelectric field;
because $E^3$ has the helicity zero, $(F^{ + R})^\dag E^3$ and $(F^{ + L})^\dag E^3$ represent the helicity-flip by $\pm1$. Similarly, the bilocal operator 
${F^{ +  \bot }}{F^{12}}$
relevant to $G_{3H}(x)$ is 
expressed by $(F^{ + R})^\dag H^3$, $(F^{ + L})^\dag H^3$ 
with $H^3= -F^{12}$ being the third component of the chromomagnetic field;
because $H^3$ has the helicity zero,  $(F^{ + R})^\dag H^3$ and 
$(F^{ + L})^\dag H^3$ also represent the helicity-flip by one unit. 
With manifest gauge invariance, we have the chromoelectric and chromomagnetic
helicity-zero contributions, $E^3$ and $H^3$, for the gluon,
so that we have the two types of contributions,
$G_{3E}$ and $G_{3H}$, for the helicity-flip by one unit 
relevant to the transverse-spin distribution.
This explains why the transverse-spin gluon distribution $G_T$
is given as the sum of the two distributions
as in (\ref{gspint}).

We discuss the nucleon spin sum rules using our results.
The usual spin sum rule expresses the total angular momentum for the longitudinally-polarized nucleon,
${J_\parallel } = \frac{1}{2}$, as the sum of the orbital angular momentum contribution $L$,
the quark spin contribution $\Delta \Sigma$,
and the gluon spin contribution $\Delta G$, and reads, using the above-mentioned helicity PDFs,
\begin{equation}
\frac{1}{2} = L + \frac{1}{2}\Delta \Sigma  + \Delta G\ ,\;\;\;\; \;\;\;\; \;\;\;\; \;\;\;\; 
\Delta \Sigma  \equiv \int {dx} \Delta q(x)\ , \;\;\;\;\;\;\;\; \;\;\;\;\;\;\;\;    \Delta G \equiv \int {dx} \Delta G(x)\ .
\label{lsr}
\end{equation}
Here and below, the summation over all quark and antiquark flavors for the quark spin contribution is implicit.
Using 
(\ref{qpdfr}),
$\Delta \Sigma$ is given as matrix element of local
operator:
$\Delta \Sigma=(\sqrt 2 /P^ + )\langle P{\rm{ }}{S_\parallel }|{\psi ^\dag }(0){\hat s^3}$
$\psi (0)|P{\rm{ }}{S_\parallel }\rangle$.
Substitution of (\ref{gspinl}) into (\ref{lsr}) gives,
\begin{equation}
\Delta G = \frac{(n^-)^2}{2}\int {dx} \frac{1}{x}\int {\frac{{d\lambda }}{{2\pi }}} {e^{i\lambda x}}\langle P{\rm{ }}{S_\parallel }|\left(F^{\alpha \beta } (0)\right)^\dag{\hat s^3}{F^{\alpha \beta }}(\lambda n)|P{\rm{ }}{S_\parallel }\rangle\ ,
\label{dg}
\end{equation}
where the factor $1/x$ in the integrand
prevents from obtaining the local operator.
The similar sum rule 
for the total angular momentum 
of the transversely-polarized nucleon, $J_T =\frac{1}{2}$, should read,
\begin{equation}
\frac{1}{2} = L_T + \frac{1}{2}\Delta_T \Sigma  + \Delta_T G\ ,\;\;\;\; \;\;\;\; \;\;\;\; \;\;\;\; 
\Delta_T \Sigma  \equiv \int {dx}  g_T(x)\ , \;\;\;\;\;\;\;\; \;\;\;\;\;\;\;\;    \Delta_T G \equiv \int {dx} G_T(x)\ .
\label{tsr}
\end{equation}
and, substituting 
(\ref{qpdfr}) and (\ref{gspinl}), 
we find, 
$\Delta_T\Sigma = (1/M) \langle P{\rm{ }}{S_ \bot }|{\psi ^\dag }(0){\hat s^ \bot }\psi (0)|P{\rm{ }}{S_ \bot }\rangle$, and 
\begin{equation}
\Delta_T G = \frac{n^-}{{2\sqrt 2 M}}\int {dx} \frac{1}{x}\int {\frac{{d\lambda }}{{2\pi }}} {e^{i\lambda x}}\langle P{\rm{ }}{S_ \bot }|\left(F^{\alpha \beta } (0)\right)^\dag{\hat s^ \bot }{F^{\alpha \beta }}(\lambda n)|P{\rm{ }}{S_ \bot }\rangle\ , 
\label{dgt}
\end{equation}
i.e.,
the result for the gluon spin contribution
is again given by the integral of the bilocal operator.
Apparently, the two matrix elements of the local operators 
for $\Delta \Sigma$ and $\Delta_T \Sigma$ are
related by the space rotation in the nucleon rest frame. 
We note that the above formulas
are obtained for the fast-moving nucleon.
For the quark spin contributions, we can immediately derive
the similar formulas in the rest frame with $\bm{P}=0$: 
$\Delta_T \Sigma$ is given by the same formula as above,
while $\Delta \Sigma=(1/M)\langle P{\rm{ }}{S_\parallel }|{\psi ^\dag }(0){\hat s^3}\psi (0)|P{\rm{ }}{S_\parallel }\rangle$.
Thus, rotation symmetry allows us to conclude
$\Delta \Sigma=\Delta_T \Sigma$.  
Now, the remaining question is whether $\Delta G$ and $\Delta_T G$ are equal or not.
This is a nontrivial question, because both $\Delta G$ and $\Delta_T G$ are given as matrix elements
of the nonlocal operators depending explicitly on a fixed light-like vector $n^\mu$,
as in  (\ref{dg}), (\ref{dgt}).

To analyze this problem, 
we treat the bilocal operators in (\ref{gspint}),
corresponding to $G_{3E}(x)$ and $G_{3H}(x)$,
based on the operator product expansion, which manifestly satisfies
Lorentz as well as rotation symmetry.
The operator product expansion for $G_{3E}(x)$ is obtained in \cite{Hatta:2012jm} recently, and
allows us to obtain the formula, 
${G_{3E}}(x) = \int_x^1 d y\frac{{\Delta G(y)}}{{y}} + \left[ {{\mbox{genuine twist-three}}} \right]$;
here, the first term corresponds to the Wandzura-Wilczek contribution
as an integral of the gluon helicity distribution $\Delta G(x)$, and the second term
denotes the genuine twist-three contributions which are expressed by certain integrals of
the three-gluon correlations 
$\sim \langle P{\rm{ }}{S_ \bot }|{F^{ +  \bot }}{F^{ +  \bot }}{F^{ +  \bot }}|P{\rm{ }}{S_ \bot }\rangle$
and the quark-gluon three-body correlations
$\sim \langle P{\rm{ }}{S_ \bot }|\bar \psi {F^{ +  \bot }}\psi |P{\rm{ }}{S_ \bot }\rangle$.
On the other hand, the bilocal operators for $G_{3H}(x)$ in (\ref{gspint}) prove to have appeared 
in the intermediate stage of the operator product expansion (of  the flavor-singlet part) of the
structure function $g_2 (x, Q^2)$, which is the twist-three structure 
function in the DIS of the transversely-polarized nucleon off the longitudinally-polarized
lepton.
Using the results in 
\cite{Kodaira:1997ig,Kodaira:1998jn,Braun:2000yi}
for the operator product expansion of the corresponding flavor-singlet part,
we find that  $G_{3H}(x)$ is given
solely by the genuine twist-three contributions
in terms of $\langle P{\rm{ }}{S_ \bot }|{F^{ +  \bot }}{F^{ +  \bot }}{F^{ +  \bot }}|P{\rm{ }}{S_ \bot }\rangle$,
$\langle P{\rm{ }}{S_ \bot }|\bar \psi {F^{ +  \bot }}\psi |P{\rm{ }}{S_ \bot }\rangle$.
It is straightforward to see that the first moment of those genuine twist-three contributions vanishes,
so that we obtain, 
$\int {dx} {G_{3E}}(x) = \Delta G$~\cite{Hatta:2012jm}, and $\int {dx} {G_{3H}}(x) = 0$~\cite{kt}.
As a result, we find,
$\Delta_T G=\Delta G$,
similarly as the quark spin contributions.
This result coincides with the gluon spin contributions 
calculated in \cite{Hatta:2012jm}
using gauge-invariant  
decomposition of the QCD angular momentum tensor into the
quark/gluon contributions (see \cite{Chen:2008ag}).

We finally mention about the orbital angular momentum contribution to
the spin sum rules.
For the longitudinally-polarized case~(\ref{lsr}), the orbital angular momentum contribution 
can be further decomposed into the well-defined quark and gluon contributions 
in a frame- and model-independent way as
$L=L_q+L_g$, but in many ways as discussed by many authors~\cite{Chen:2008ag,Leader:2013jra}.
On the other hand, for the transversely-polarized case~(\ref{tsr}), the corresponding 
quark/gluon contributions $\left (L_T\right)_{q,g}$, such that 
$L_T=\left (L_T\right)_q+\left (L_T\right)_g$,
receive the terms
${\bar C_{q,g}}P^3/[2({P^0} + M)]$, respectively, which are frame dependent~\cite{Hatta:2012jm};
here $\bar C_{q,g}$ arise in matrix element of
the QCD angular momentum tensor ${\cal M}^{\lambda \mu\nu} = x^\mu T^{\lambda \nu} -x^\nu T^{\lambda \mu}$,
using 
the well-known parameterization of the 
off-forward matrix element of the  (Belinfante-improved) energy-momentum tensor of quarks/gluons,
\begin{equation}
\langle P'S'|T_{q,g}^{\mu\nu}|PS\rangle = \bar{u}(P',S') \!\!\! \left[ \! A_{q,g} \gamma^{(\mu}\bar{P}^{\nu)}
\! +\!B_{q,g}\frac{\bar{P}^{(\mu}i\sigma^{\nu)\alpha}\Delta_\alpha}{2M} 
 \!  +\! C_{q,g}\frac{\Delta^\mu\Delta^\nu -g^{\mu\nu}\Delta^2}{M} \!+\! \bar{C}_{q,g}Mg^{\mu\nu}\! \right]\!\!\! u(P,S)\ ,
\label{eq8}
\end{equation}
with $\bar{P}^\mu=\frac{1}{2}(P^\mu+P'^\mu)$, $\Delta^\mu = P'^\mu-P^\mu$.
The frame-dependent terms disappear in $L_T$ because $\bar C_q+\bar C_g=0$.
Thus, we do not have a well-defined decomposition of the orbital angular momentum 
contribution $L_T$ into the quark and gluon contributions for the transversely-polarized case~(\ref{tsr}).

To summarize,
we have given a QCD definition of the transverse-spin gluon distribution function
based on the spin-operator representation for bilocal operator definitions
of the PDFs.
This definition has the structure of the quantum mechanical expectation value
in the spin space,
$\int {\frac{{d\lambda}}{{2\pi }}} {e^{i\lambda x}}\langle P{\rm{ }}S| \Phi^\dagger(0)\hat{O}
\Phi (\lambda n)|P{\rm{ }}S\rangle$ with 
$\hat{O}=1, \hat{\bm{s}}$,
in a unified form 
for the quark ($\Phi  = \psi$) and gluon ($\Phi  = F^{\mu \nu }$)
distribution functions.
For both quark and gluon cases,
$\hat{O}=1$ leads to the density PDFs,
$\hat{O}=\hat{s}^3$ leads to the helicity PDFs,
and $\hat{O}=\hat{s}^\bot$ leads to the transverse-spin PDFs
associated with helicity-flip by one unit.
We have shown that the new transverse-spin distribution function, $G_T(x)$,
is given as the sum of the two twist-three gluon distribution
functions, $G_{3E}(x)$ and $G_{3H}(x)$,
which arise in the most general decomposition of the gluonic correlator, (\ref{ff}).
The operator product expansions relevant to these new gluon distributions are
available, and allow us to show that the first moments of $\Delta G(x)$ and
$G_T(x)$, which respectively represent the gluon spin contributions to the nucleon spin 
for the longitudinally- and transversely-polarized cases,
are equal. The operator product expansion can be exploited also
to analyze the higher moments, $\int dx x^{n-1} G_T (x)$~\cite{kt}.
Finally, to seek hard processes which allow direct access
to $G_T(x)$ is an interesting future problem.

\section*{Acknowledgements}
I thank Y.~ Hatta and S.~Yoshida for discussions and collaborations.
This work is supported in part by the Grant-in-Aid for Scientific Research
 (Nos.~24540284, 25610058 and 26287040).

\end{document}